\documentclass[twocolumn,amsmath,amssymb,10pt,superscriptaddress,a4paper,letterpaper,fleqn]{revtex4-1}
\usepackage{amssymb}
\usepackage{epsfig}
\usepackage{graphicx}
\usepackage{dcolumn}
\usepackage{array}
\usepackage{bm}
\usepackage{fancyheadings}
\usepackage{longtable}
\usepackage{multirow}
\usepackage{float}
\usepackage{afterpage}
\usepackage{color}
\setlongtables
\usepackage[breaklinks=true,linkbordercolor={1 1 1}]{hyperref}
\newcommand{\gras}[1]{\boldsymbol{#1}}

\begin{document}
\setcounter{page}{1}
\title{
     \qquad \\ \qquad \\ \qquad \\  \qquad \\  \qquad \\ \qquad \\ 
 Quantification of Uncertainties in Nuclear Density Functional theory
}

\author{N. Schunck}
\email[Corresponding author: ]{schunck1@llnl.gov}
\affiliation{Physics Division, Lawrence Livermore National Laboratory,
Livermore, CA, 94551, USA}
\author{J.D. McDonnell}
\affiliation{Physics Division, Lawrence Livermore National Laboratory,
Livermore, CA, 94551, USA}
\author{D. Higdon}
\affiliation{Los Alamos National Laboratory, Los Alamos, New Mexico 87545, USA}
\author{J. Sarich}
\affiliation{Mathematics and Computer Science Division, Argonne National
Laboratory, Argonne, IL 60439, USA}
\author{S. Wild}
\affiliation{Mathematics and Computer Science Division, Argonne National
Laboratory, Argonne, IL 60439, USA}

\date{\today}
   \received{XX May 2014; revised received XX August 2014; accepted XX September 2014}

\begin{abstract}{
Reliable predictions of nuclear properties are needed as much to answer
fundamental science questions as in applications such as reactor physics or
data evaluation. Nuclear density functional theory is currently the only
microscopic, global approach to nuclear structure that is applicable throughout
the nuclear chart. In the past few years, a lot of effort has been devoted to
setting up a general methodology to assess theoretical uncertainties in nuclear
DFT calculations. In this paper, we summarize some of the recent progress in
this direction. Most of the new material discussed here will be be published
in separate articles.
}
\end{abstract}
\maketitle

\lhead{Quantification of Uncertainties $\dots$}    
\chead{NUCLEAR DATA SHEETS}                       
\rhead{N. Schunck \textit{et al.}}              
\lfoot{}
\rfoot{}
\renewcommand{\footrulewidth}{0.4pt}



\section{INTRODUCTION}

The rapid development of leadership class computing facilities throughout the
world, accompanied by targeted programs from funding agencies to foster the
use of high-performance computing methods in science, have opened new
opportunities in theoretical nuclear structure~\cite{bogner2013}. It has become
possible to address important questions of nuclear science using microscopic
approaches to structure and reaction rooted in the knowledge of effective
nuclear forces and standard methods of quantum mechanics. Recent examples
include the explanation of the anomalously long half-life of $^{14}$C isotope
used in carbon-dating~\cite{maris2011}, predictions of neutrino-nucleus
currents relevant to physics beyond the standard model~\cite{lovato2014}, or
of light ion fusion reactions relevant to the National Ignition
Facility~\cite{navratil2012}, to name but a few.

In parallel, there has been an increasing need for accurate and precise data,
whether from measurements or simulations, in areas as diverse as nuclear
astrophysics~\cite{kaeppeler2011,langanke2003}, reactor
physics~\cite{ceresio2012} or data evaluation~\cite{herman2011}. In the past,
the cost of using standard methods of statistics to estimate theoretical
uncertainties in such microscopic approaches was often prohibitive, but this
limitation has slowly been disappearing.

Among the few microscopic theories of nuclear structure, density functional
theory (DFT) plays a special role, since it is the only one to be applicable
across the entire nuclear chart, from the lightest to the heaviest elements.
Therefore, DFT is the tool of choice to study phenomena such as nuclear
fission~\cite{younes2011} or superheavy element predictions~\cite{hofmann2000},
but has also seen applications in tests of fundamental
symmetries~\cite{dobaczewski2005,ban2010} or the search for neutrino-less
double beta-decay~\cite{rodriguez2010}.

In this proceeding, we briefly present some of the challenges and methodologies
used in nuclear DFT to estimate theoretical uncertainties. This topic is
covered in greater details in an invited contribution to a Focus Issue of the
Journal of Physics G: Nuclear and Particle Physics on ``Enhancing the
Interaction Between Nuclear Experiment and Theory Through Information and
Statistics ''~\cite{schunck2014,wild2014,higdon2014}. In section \ref{sec:dft}, 
we recall the main components of nuclear DFT. In section \ref{sec:uq}, we 
summarize some of the recent results in uncertainty quantification and error 
propagation, before we conclude in section \ref{sec:conclusions}.


\section{NUCLEAR DENSITY FUNCTIONAL THEORY}
\label{sec:dft}

Density functional theory (DFT) is a general approach to the quantum many-body
system. It is based on a series of theorems by Kohn and Sham, who have shown
that it is theoretically possible to find the exact ground-state energy of a
system of $N$ interacting electrons by solving a system of equations
characteristic of an independent particle system~\cite{hohenberg1964,kohn1965}.
This existence theorem was later extended to the context of nuclear
physics~\cite{messud2009}. Nuclear DFT is a reformulation of the traditional
self-consistent mean-field (SCMF) theory of nuclear structure, which has been
very successful in predicting a broad range of nuclear properties.

The essential component of both the SCMF theory and nuclear DFT is the energy
density functional (EDF), which encapsulates all information about the system
(in principle). The EDF is a functional of the density of neutrons and protons, as
well as of the pairing density~\cite{perlinska2004}. In nuclear DFT, the EDF
is treated at the Hartree-Fock-Bogoliubov (HFB) approximation~\cite{ring1980};
in the SCMF, the EDF is often related to an underlying two-body Hamiltonian,
and the HFB approximation may be only the first step of a series of
calculations~\cite{bender2003}. In any case, the EDF is characterized by a
number of coupling constants which are not given by any underlying theory and
must therefore be adjusted to some experimental data.

One must emphasize that the Kohn-Sham theorem is only an existence theorem:
there is no magic recipe to determine the one EDF that will give the exact
energy of the nucleus. In addition, in-medium nuclear forces are poorly known
and should in principle be derived from quantum chromodynamics. This is in
contrast to electronic DFT, where the Coulomb force is known exactly. For these
reasons, one should, therefore, consider nuclear DFT (and the SCMF) as
inherently imperfect models of the nucleus: this is the first, major source of
errors in DFT, which we will refer to as ``model errors''. Let us denote by
$\gras{x} = (x_{1}, \dots, x_{n_{x}})$ the parameters of the EDF, aka the
model. Typically, $n_{x} \approx 10-20$. These parameters will be fitted on
some $n_{d}$ data points $y_{i}$. There could be different types of data:
atomic masses, r.m.s. charge radii, mass differences, excitation energies of
isomers, etc. It is clear that, given a specific EDF, the choice of the
experimental data will impact the overall predictive power of DFT: this is the
second source of errors in DFT, which we will label ``fitting errors''.
Finally, there is a third source of errors, ``implementation errors'', caused
by the need to solve the DFT equations numerically. These various sources of
uncertainties are discussed in more details in~\cite{schunck2014}. In this
proceeding, we focus only on selected aspects of fitting errors.


\section{QUANTIFYING AND PROPAGATING ERRORS IN NUCLEAR DFT}
\label{sec:uq}

As already mentioned above, we will only discuss uncertainties pertaining to
the determination of model parameters. We thus assume we have an energy density
functional, which is characterized by the $n_{x}$ unknown parameters
$\gras{x}$. We are trying to determine the best way to obtain an optimal set of
parameters $\gras{x}$ and, in the same time, to quantify the uncertainties
associated with this procedure.

We recall that there is a very large amount of experimental data
that potentially could be used to fit the few parameters of an EDF. However, different data
types may have very different impacts on specific model parameters. For
example, it was pointed out using a singular value decomposition analysis that
only a few of the eight parameters of a standard Skyrme EDF are really relevant
to reproduce nuclear masses~\cite{bertsch2005} or single-particle
energies~\cite{kortelainen2008}. In order to constrain every coupling constant
of the EDF, it thus appears necessary to introduce different types of data. In
practice, the determination of EDF parameters is thus made by minimizing the
composite $\chi^{2}$ function
\begin{equation}
\chi^{2}(\gras{x}) = \frac{1}{n_{d} - n_{x}} \sum_{t=1}^{n_{T}} \sum_{j=1}^{n_{t}}
\left( \frac{y_{tj}(\gras{x}) - d_{tj}}{\sigma_{t}} \right)^{2},
\label{eq:chi2}
\end{equation}
with $n_{T}$ the number of different data types, $n_{t}$ the number of data
points for type $t$, and $n_{d} = \sum_{t} n_{t}$ the total number of data
points over all types. The calculated value of data point number $j$ of type
$t$ is denoted by $y_{tj}$, with $d_{tj}$ the corresponding experimental
value. Because there are different types of data, relative distances must be
properly normalized by the quantity $\sigma_{t}$, which represents an estimate
of the theoretical error on data type $t$. This strategy was followed in a
series of paper by the UNEDF
collaboration~\cite{kortelainen2014,kortelainen2012,kortelainen2010}.

The minimization of the $\chi^{2}$ function gives access to the ``optimal''
parametrization of the EDF. Obviously, one should bear in mind that this
optimal choice is strongly dependent on (i) the choice of the types $t$ of
experimental data, (ii) the number of data points for each type $t$, (iii) the
weight $\sigma_{t}$ chosen for each type. In addition, the quality of the
optimization is contingent of the algorithm used and depends on the starting
point. Bearing in mind these caveats, it is possible to estimate the covariance
matrix by assuming normally distributed errors and approximate linear
variations of the $\chi^{2}$ function under variations of model
parameters~\cite{dobaczewski2014}. This approximation has often been used to
propagate model
errors~\cite{reinhard2013,piekarewicz2012,reinhard2013-a,kortelainen2013}.

Very recently, alternative approaches to uncertainty quantification based on
Bayesian statistics have been investigated for semi-microscopic nuclear mass 
models based \cite{goriely2014}. In the context of nuclear DFT, such approaches 
are appealing since they treat model parameters as intrinsically random 
variables, the true value of which can not be known with certainty. This 
perspective is particularly adapted to nuclear structure theory, since the 
nuclear many-problem is unsolvable exactly: only approximations are available
(DFT is one of them), and, therefore, uncertainties are unavoidable and should
be quantified. Mathematical details on how posterior distributions can be 
generated in the context of nuclear DFT are discussed in details in 
\cite{higdon2014}; a paper currently being finalized by our group also uses 
Bayesian posteriors to analyze theoretical uncertainties for the prediction of 
neutron-drip lines and fission barriers in actinides \cite{mcdonnell2014}. 

\begin{figure}[!htb]
\includegraphics[width=0.97\columnwidth]{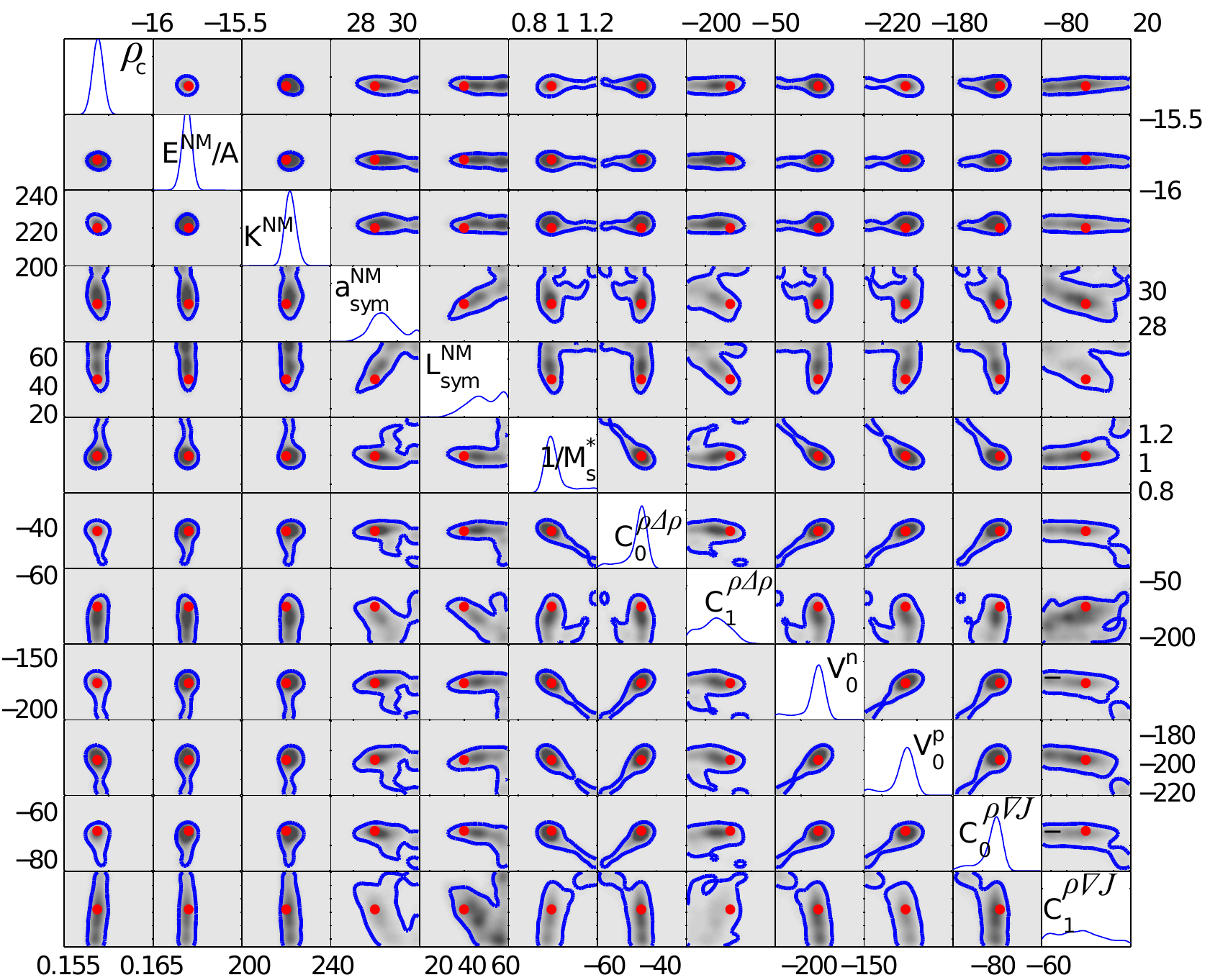}
\caption{(Color online) Univariate and bivariate marginal estimates of the 
posterior distribution for the 12-dimensional DFT parameter vector $\gras{x}$ 
of the UNEDF1 parametrization. The blue line encloses an estimated 95\% region.
.}
\label{fig:uq}
\end{figure}

We show in Fig.~\ref{fig:uq} one of the first examples of a Bayesian posterior 
distribution corresponding to the UNEDF1 $\chi^{2}$ function of 
\cite{kortelainen2010}. The red dots correspond to the UNEDF1 solution itself. 
Obtaining such distributions requires first to set up intervals of variations  
$[x_{i}^{\text{min}}, x_{i}^{\text{max}}]$ for each of the DFT parameters. 
These intervals define a 12-d hypercube in parameter space from which the prior 
distribution is sampled. In the case shown in Fig.~\ref{fig:uq}, the interval 
for each parameter $x_{i}$ was defined as 
$[x^{*}_{i} - 3\sigma_{i}, x_{i}^{*}+3\sigma_{i}]$, with $x_{i}^{*}$ the 
UNEDF1 value and $\sigma_{i}$ its standard deviation; see 
\cite{kortelainen2012}. Since this work is still exploratory, we chose a 
uniform prior distribution. Because of the significant cost of running the DFT 
calculation of the UNEDF1 $\chi_{2}$ (about 800 cores for 15 minutes), the 
posterior was extracted from a response function constructed using Gaussian 
Process techniques; see \cite{higdon2014} for a full description of the 
method. The response function was based on 200 DFT calculations of the 
$\chi_{2}$ sampling the 12-d hypercube. 

\begin{figure}[!htb]
\includegraphics[width=0.97\columnwidth]{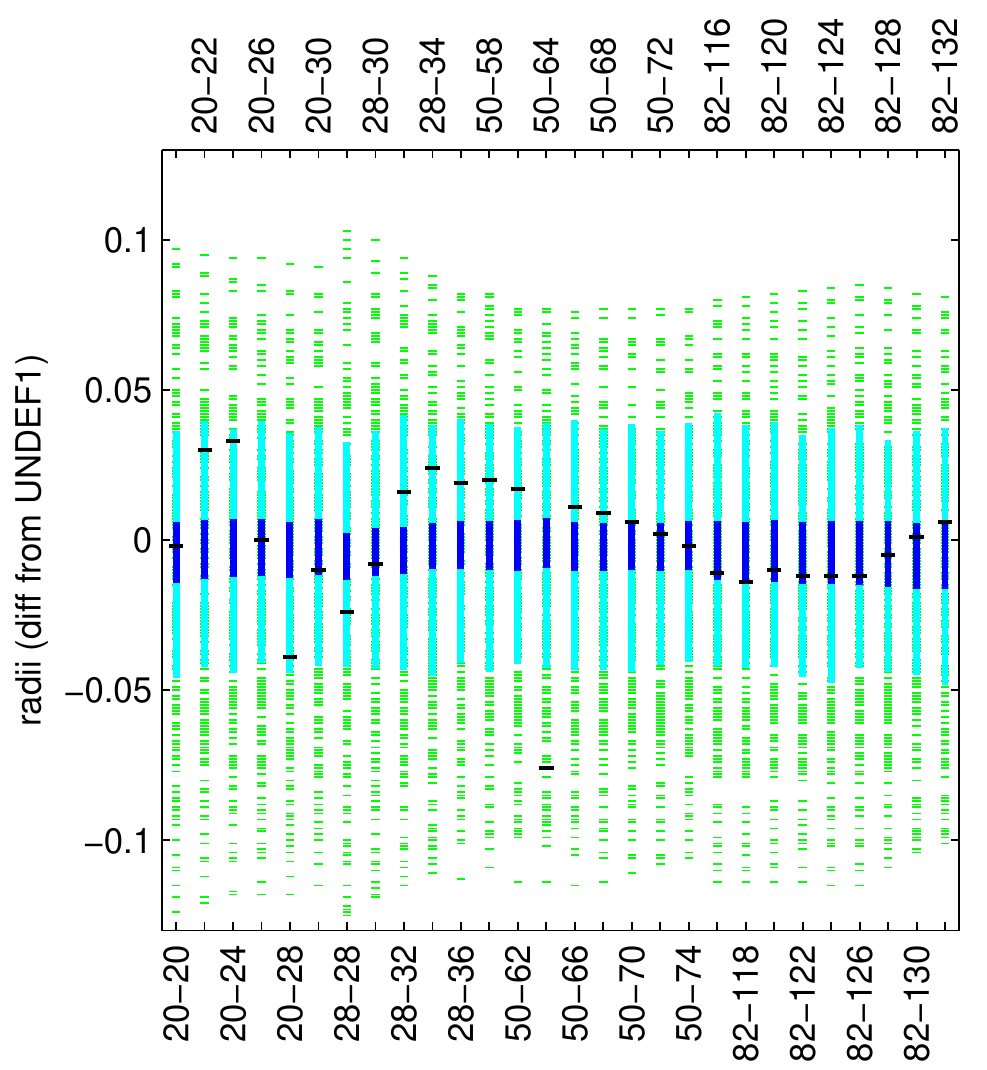}
\caption{(Color online) Estimates of theoretical uncertainties for proton 
radii in 27 spherical Ca, Ni, Sn and Pb isotopes relative to the UNEDF1 
parametrization of the Skyrme functional. Nuclei are labeled Z\_N. Black 
marks are experimental values, the dark blue band the 90\% prediction 
interval of the model alone, the light blue band the 90\% prediction 
interval when also including fitting errors of the model.
}
\label{fig:radii}
\end{figure}

Once the posterior distribution for the DFT model parameters is known, we can 
estimate theoretical error bars of an observable $\mathcal{O}$ by computing 
it for a sample $\mathcal{S} = (\gras{x}_{1}, \dots, \gras{x}_{s})$ of DFT 
parameter sets drawn from the posterior. As mentioned earlier, we will 
present the results of such a strategy for neutron drip lines and fission 
barriers elsewhere \cite{mcdonnell2014}. In this proceeding, we illustrate 
the approach with the preliminary example of the proton r.m.s. radii in 27 
spherical nuclei used in the fit of the UNEDF family of functionals. 
Fig.~\ref{fig:radii} shows the estimate of theoretical errors obtained 
from the Bayesian analysis relative to the UNEDF1 values. Black marks are 
the experimental values of the radius. The dark blue band shows the 90\% 
prediction uncertainty (including emulator error). The light blue band 
also includes the fitting error, i.e., the discrepancy between the actual 
experimental data and the DFT calculation -- assumed to follow a normal 
distribution. The outlier at Z=50, N=64 hints at systematic errors, i.e. 
the inability of the model to reproduce the data, irrespective of how 
model parameters are fitted.


\section{CONCLUSIONS}
\label{sec:conclusions}

In many important research areas and contemporary applications of nuclear
science, nuclear density functional theory represents the only microscopic
model of structure and reactions available. In this proceeding, we have briefly
summarized some of the challenges and recent results in identifying and
quantifying theoretical uncertainties inherent to nuclear DFT. In particular,
we have emphasized the widespread use of covariance analysis and the first
applications of Bayesian statistics in DFT. With the constant development of
supercomputers, such methods will most likely gain in popularity and could be
applied, e.g., to practical applications such as the quantification of errors for
fission product yields in neutron-induced fission.


{\it Acknowledgements:} This work was partly performed under the auspices of
the U.S.\ Department of Energy by Lawrence Livermore National Laboratory under
Contract DE-AC52-07NA27344. It was supported by the SciDAC activity within the
U.S.\ Department of Energy, Office of Science, Advanced Scientific Computing
Research program under contract number DE-AC02-06CH11357. Computational
resources were provided through an INCITE award ``Computational Nuclear
Structure'' by the National Center for Computational Sciences (NCCS) and
National Institute for Computational Sciences (NICS) at Oak Ridge National
Laboratory, through an award by the Livermore Computing Resource Center at
Lawrence Livermore National Laboratory, and through an award by the Laboratory
Computing Resource Center at Argonne National Laboratory.
\!\!\!\!\!\!\!\!


\end{document}